\pdfoutput=1
\documentclass[aps,pre,superscriptaddress,twocolumn,nofootinbib]{revtex4-2}

\usepackage[dvipsnames]{xcolor}
\usepackage{amsmath,amssymb}
\usepackage{graphicx}
\usepackage{grffile} 
\usepackage{footmisc}
\usepackage{bbold,soul}
\usepackage{lipsum}
\usepackage{enumitem}

\usepackage[colorlinks=true,bookmarks=false,citecolor=blue,linkcolor=red,hyperfootnotes=true,urlcolor=blue]{hyperref}

\makeatletter
\newcommand\footnoteref[1]{\protected@xdef\@thefnmark{\ref{#1}}\@footnotemark}
\makeatother

\graphicspath{{./}{./figures/}}

\newcommand{\pd}{P_\downarrow}
\newcommand{\pu}{P_\uparrow}

\newcommand{\bc}{\beta_\mathrm{c}}

\newcommand{\To}{T_\mathrm{o}}
\newcommand{\Td}{T_\mathrm{d}}

\newcommand{\Eth}{E_\mathrm{th}}
\newcommand{\Ea}{E_\mathrm{a}}
\newcommand{\Er}{E_\mathrm{ridge}}

\begin{document}
\title{Competition between Barrier- and Entropy-Driven Activation in Glasses
}
\author{Matthew R. Carbone}
\affiliation{Computational Science Initiative, Brookhaven National Laboratory, Upton, New York 11973, USA}
\author{Marco Baity-Jesi}
\email[]{marco.baityjesi@eawag.ch}
\affiliation{Eawag (ETH), \"Uberlandstrasse 133, CH-8600 D\"ubendorf, Switzerland}

\date{\today}

\begin{abstract}
In simplified models of glasses we clarify the existence of two different kinds of activated dynamics, which coexist, with one of the two dominating over the other. One is the energy barrier hopping that is typically used to picture activation, and the other one, which we call entropic activation, is driven by the scarcity of convenient directions. When entropic activation dominates, the height of the energy barriers is no longer the decisive to describe the system's slowdown.
In our analysis, dominance of one mechanism over the other depends on the shape of the density of states and temperature. We also find that at low temperatures a phase transition between the two kinds of activation can occur. Our framework can be used to harmonize the facilitation and thermodynamic pictures of the slowdown of glasses.
\end{abstract}

\maketitle


\paragraph*{Introduction.}
Glasses are inherently slow systems. This slowness can be captured by mean-field (MF) theory, which recently brought to a series of breakthroughs that allowed for a deep understanding of the reasons underlying their sluggishness~\cite{charbonneau:14,maimbourg:16,charbonneau:17}. However, MF theory predicts divergences of the relaxation time that do not occur in real systems, because it does not capture relaxation mechanisms that only appear in low-dimensions. These are generically called \emph{activation}, and they are most often pictured as the hopping of energy barriers~\cite{doliwa:03c}: since in MF the barriers diverge with the system size $N$, a simple argument is that activation in MF cannot occur because barriers cannot be hopped~\cite{arceri:20}.


However, activation can be studied in the MF limit by specifying to finite-$N$ systems and very long times~\cite{crisanti:00,crisanti:00b}. Several works focused on comparing the dynamics of simple models, such as the Random Energy Model (REM)~\cite{derrida:80}, to the Trap Model (TM)~\cite{dyre:87,bouchaud:92,bouchaud:95}, to work out whether the barrier hopping dynamics can be assimilated to jumping between traps with a fixed threshold energy~\cite{benarous:02,benarous:08,gayrard:16,gayrard:16b,cerny:17,gayrard:18,gayrard:19,baityjesi:18,baityjesi:18c}.
Other works studied the saddles connecting minima~\cite{ros:19,ros:19b,ros:20,ros:21}, extensions of the Franz-Parisi potential~\cite{franz:20}, or path-integral approaches to study the dynamics between different minima~\cite{rizzo:21}.

Comparisons with the TM were also performed in models with a trivial landscape, such as the Step~\cite{cammarota:15} or Funnel models~\cite{carbone:20}, where it was shown that entropic effects can lead to Trap-like activation, if instead of considering basins in phase space we construct them dynamically. Indications of entropic effects in long-time dynamics was also found in less idealized systems, such as the $p$-spin model~\cite{stariolo:19,stariolo:20}, finite-connectivity Step models~\cite{tapias:20}, or even 3$D$ Lennard-Jones mixtures at $\Td<T<\To$~\cite{baityjesi:21}.
In these works, however, the underlying framework is either that activation only exists as barrier hopping~\cite{bouchaud:92,doliwa:03c}; either that it is an entropic effect that can be assimilated to barrier hopping~\cite{cammarota:15,carbone:20}; or it is a transient behavior which eventually turns into hopping~\cite{crisanti:00,tapias:20}. This is hard to reconcile with other pictures of the dynamical arrest of glasses, such as the facilitation picture, that argue that the landscape (energy barriers) is not crucial to explain the slowness of glasses, which should instead be attributed to kinetic constraints~\cite{chandler:10,royall:20}.

Here, we clarify the nature of these entropic effects, showing how under the right lens they can be used to unify the landscape with the facilitation pictures.
Specifically, we take a paradigmatic model of glasses, the Random Energy Model (REM)~\cite{derrida:80}, and show that both barrier- and entropy-driven activation mechanisms coexist. When barrier-driven activation dominates, the system's slowness is driven by the energy barrier separating basins, while when entropic activation dominates, the height of the barrier becomes unimportant, and the slowness is instead driven by the scarcity of convenient directions. 
In our analysis, the dominance of one mechanism or the other depends on the density of states $\rho(E),$ and on the temperature $T\equiv\beta^{-1}$, which is the control parameter of a non-equilibrium phase transition between the two different activated regimes.

\paragraph*{Models.}
In the REM, we have a spin system with $N$ spins, $s_i=\pm1$. The energy of a spin configuration is independent of the configuration itself, and is instead drawn from a probability distribution $\rho(E).$ Two choices of $\rho(E)$ that are used in literature. The initial formulation of the REM used a Gaussian energy distribution,
$\rho_{g}(E) = (\sqrt{2\pi N})^{-1/2} \exp\{-E^2/2N\},$
while more recent efforts also considered the Exponential REM (EREM), which differs in that it has an exponential energy distribution~\cite{baityjesi:18},
\begin{equation} \label{rho_erem}
    \rho_{e}(E) = \bc e^{\bc E}\Theta(-E)\,.
\end{equation}
where we set $\bc=1$.
These models have a transition from a paramagnetic to a glassy phase at inverse temperatures $\beta_\mathrm{g}=\sqrt{2\ln{2}}$ and $\beta_\mathrm{e}=\bc$,
respectively.

\paragraph*{Threshold and attractor energies.}
In both models, we can define a threshold energy, analogous to that of the $p$-spin model~\cite{castellani:05}, above which minima typically do not appear. We define it following Ref.~\onlinecite{baityjesi:18}, by calculating the energy for which the probability $\pd(E)$ of finding a lower-energy configuration is $1/N.$ To leading order, the threshold energies of REM and EREM are\footnote{In our REM simulations we calculated the threshold energy numerically, to avoid the preasymptotic corrections, which are large (around 13\% in the largest sizes).}
\begin{align}\label{eq:threshold}
    \Eth^g =& -\sqrt{2N\ln N}\,,\\
    \Eth^e =& -\frac{1}{\bc}\ln N\,.
\end{align}
After long enough times, the system will typically find itself in a configuration with extensively deep energy~\cite{baityjesi:18,hartarsky:19}, which we call trap. In order to transition from one trap/basin to another, the system will need to climb to the threshold. Energetic activated dynamics would be mainly driven by the jumps among these energy barriers, in a manner that is analogous to what happens in the Trap Model. This was indeed found to be the case, by looking at the limiting values of the aging functions and comparing them with the predictions of the TM~\cite{gayrard:18,baityjesi:18}.

However, motivated by the study of entropic effects in activated dynamics, we can also define another characteristic energy, which stems from toy models which represent a purely entropic kind of activation~\cite{barrat:95,carbone:20}. This energy, which we call the \emph{attractor energy},\footnote{It was already used in Ref.~\onlinecite{cammarota:15,carbone:20}, and called threshold energy, in a context where an energetic threshold energy could not be defined. However, as we see here, this energy is different from the threshold energy, so we decide to redefine the terminology.}
is defined as the energy at which the probability $\pu(E)$ of increasing the energy at the next step is equivalent to that of decreasing it: $\pu(\Ea)\equiv\pd(\Ea)$, with $\pu(E)>\pd(E)~\forall E<\Ea$. In words, due to entropic effects, the dynamics is \emph{attracted} towards the energy $\Ea$ even when $E<\Ea.$

We obtain the attractor energies for the REM and EREM by assuming Metropolis dynamics,
\begin{align}\label{eq:attractor}
    \Ea^g =& -\frac{N\beta}{2}\,,\\
    \Ea^e =& \frac1{\beta-\bc}\ln\left(\frac{2\bc-\beta}{\bc}\right)\,,~~~\bc<\beta<2\bc\,.
\end{align}
Contrary to the threshold energy, the attractor energy is well-defined even in systems with a single energy basin, such as the Step or the Funnel model. In these systems it was shown that, despite the absence of barrier-driven activation, the dynamics is a renewal process, provided that one identifies the traps dynamically, as all the configurations visited while $E<\Ea$ ~\cite{cammarota:15,carbone:20}.
With this construction, the relationship between aging functions and trapping time distributions is exactly the one predicted by the TM~\cite{cammarota:15}, and the typical time scales grow as $e^N$~\cite{carbone:20}. In other words, $\Ea$ identifies an activated dynamics which is \emph{not} driven by the height of the energy barriers (as there are none), but which shares most of the signatures of barrier-driven activation.

We can therefore try to understand the interplay between these two mechanisms. Since $\Eth$ is the minimal height at which the system must go in order to leave a trap, the entropic mechanism is not expected to play a role when $\Ea<\Eth$. By comparing Eqs.~\eqref{eq:threshold} with Eqs.~\eqref{eq:attractor}, we see that for sufficiently large sizes in the REM $\Ea(\beta)<\Eth~\forall\beta$. Therefore, we expect that activation in the REM is purely barrier-driven.

On another hand, in the EREM we have different behaviors depending on the value of $\beta$. 
When $\beta>2\bc$, the attractor energy is lower than the threshold, so the slow dynamics should be driven by the barrier heights. When instead $\bc<\beta<2\bc$ (excluding $\beta$ close to $2\bc$ which shrinks with increasing $N$), we have $\Ea>\Eth$. This indicates that even when the system manages escaping a trap, reaching $\Eth$, it will keep going up in energy, towards $\Ea$. In other words, the height of the barrier is not that important. The reason of this is that, in this regime, even though there are directions in phase space which would decrease the energy, these are too rare, and the system would rather increase its energy than invest time looking for a descending direction.

Since activated dynamics is relevant in the limit of large but finite $N$, we can work out the transition inverse temperatures $\beta_*$ by setting $\Ea(\beta_*,N)=\Eth(N)$. This gives us the $N$-dependent transitions:
\begin{align}\label{eq:transition-beta}
    \beta_*^g &=2\sqrt{\frac{2\ln N}{N}} \stackrel{N\to\infty}{\longrightarrow} 0\,, \\
    \beta_*^e &= \bc \frac{2\ln N + W(-\frac{\ln N}{N})}{\ln N}
    \stackrel{N\to\infty}{\longrightarrow} 2\bc\,,
\end{align}
where $W(z)$ is the Lambert function.\footnote{Note that $W(0) = 0.$}
In the Gaussian REM the entropic phase disappears for increasing $N$.\footnote{
In the REM, $\beta_*<\beta_c~\forall N\geq16$.
}
In the EREM, instead, the transition stays at a finite temperature, and we have entropy-driven activation at low $\beta$, and barrier-driven activation at high $\beta$.

\paragraph*{Ridge energy and phase transition.}
We run Monte Carlo simulations (details in App.~\ref{app:numerical}) to verify this transition from an barrier-driven phase at high $\beta$, to an entropy-driven phase at lower $\beta$. Since REM and EREM do not allow for exact simulations at large system sizes, for sizes $N>24$ we rely on modified dynamics, where every time that a new configuration is visited, all its neighbors, except the last-visited configuration, are drawn anew. We call this memory-1 dynamics, and elaborate more on it in App.~\ref{app:memory1}.

We measure the ridge energy, $\Er$, defined as the highest-reached energy in each basin transition, i.e. in each of the time intervals during which $E(t)>\Eth$.\footnote{To ensure that we only measure transitions between different traps, with memory dynamics we measure the inherent structure before and after the transition, and keep the transition only if they are different.} With barrier-driven activation, we expect that $\Er(\beta;N)$ will stay close to $\Eth(N)$, while for entropy-driven activation it will overshoot to higher values. 
\begin{figure}[tb]
    \centering
    \includegraphics[width=\columnwidth]{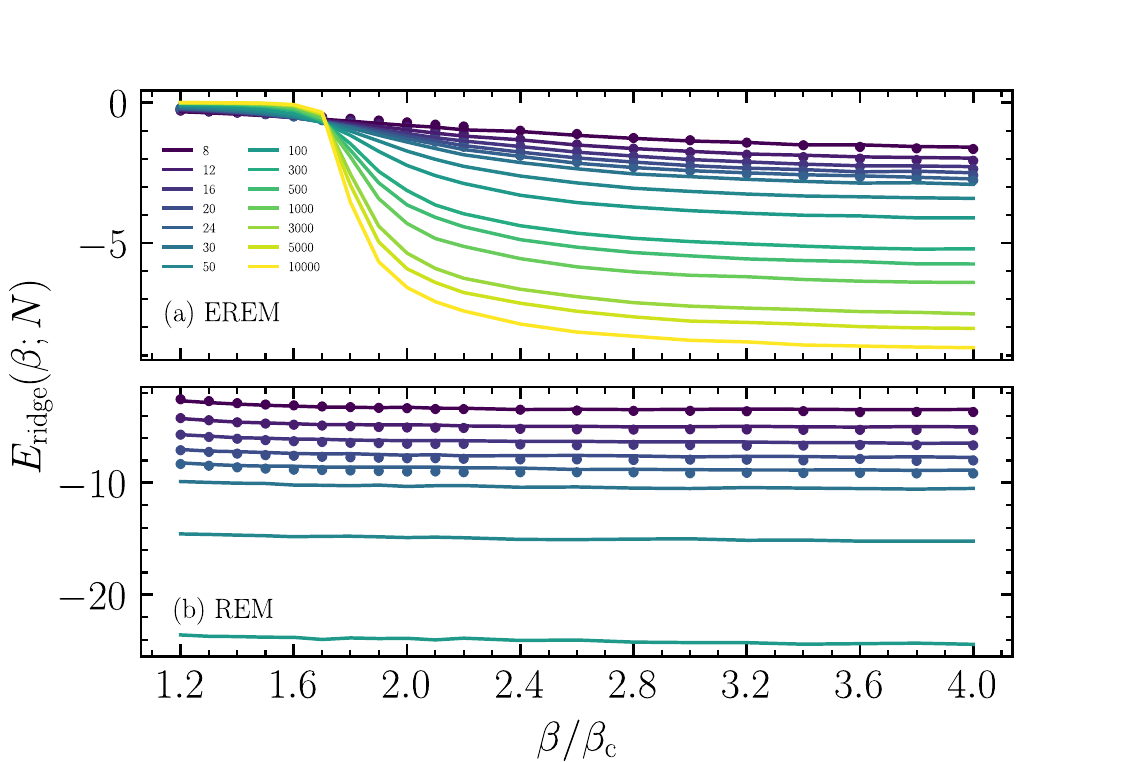}
    \caption{Median $E_\mathrm{ridge}(\beta; N)$ in the EREM (a) and REM (b). Different lines stand for different system sizes.
    Here and in Fig.~\ref{fig:scalings}, results from memory-1
    dynamics are shown in solid lines, and results from full-memory dynamics ($N<30$) are shown as markers.}
    \label{fig:ridges-beta}
\end{figure}
In Fig.~\ref{fig:ridges-beta} we plot the median $\Er$ as a function of $\beta$, for EREM and REM. While in the REM the ridge energies decrease steadily $\forall\beta$, in the EREM we see two distinct entropy-driven (low $\beta$) and barrier-driven (high $\beta$) phases.

\begin{figure}[tb]
    \centering
    \includegraphics[width=\columnwidth]{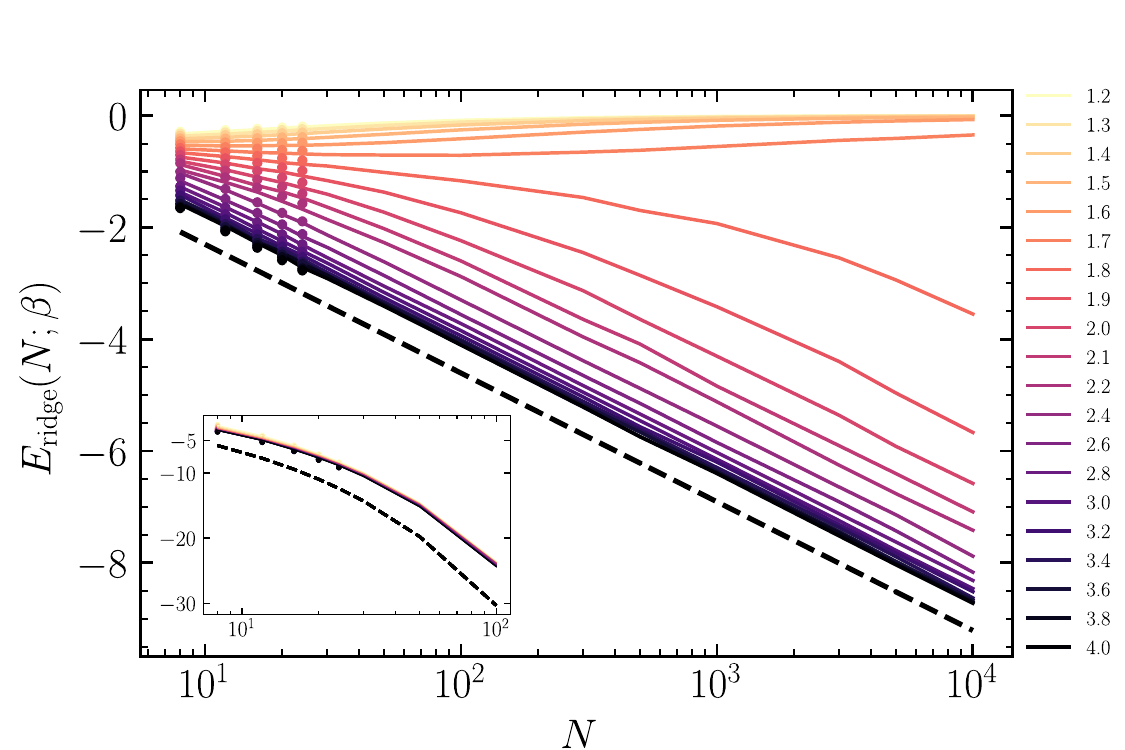}
    \caption{Median $\Er$ as a function of $N$ in the EREM and REM (inset). 
    Different lines correspond to different values of $\beta$.
    The black dashed line represents $\Eth(N)$ for each model, according to Eq.~\eqref{eq:threshold}.}
    \label{fig:scalings}
\end{figure}

As we show in Fig.~\ref{fig:scalings}, the ridge energy scales with $N$ in the way we expect. In the EREM at low $\beta$ we are in the entropic phase, and $\Er\sim-1\sim\Ea$. In the barrier-driven phase, instead, $\Er\sim-\ln(N)\sim\Eth$. In the REM, where we only have barrier-driven activation, the ridge energy follows $\Eth$ at every $\beta$.

\begin{figure}[tb]
    \centering
    \includegraphics[width=\columnwidth]{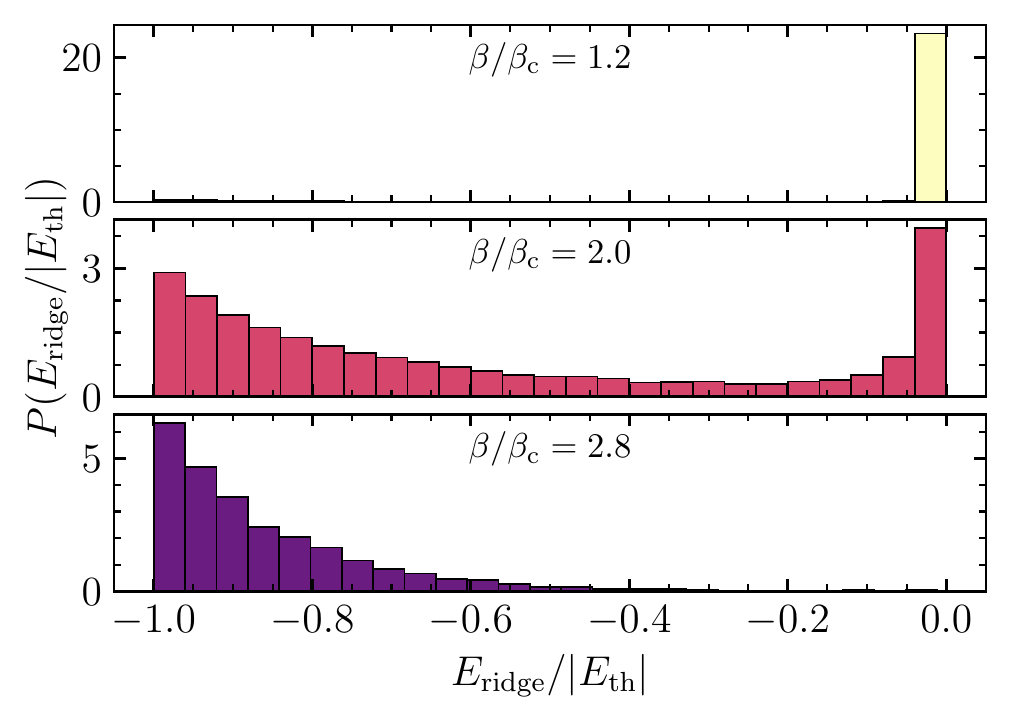}
    \caption{The density of ridge energies in the EREM model, using memory-1 dynamics, for $N=10000$. 
    }
    \label{fig:densities}
\end{figure}

Through the lens of the median $\Er,$ the transition appears at $\beta$ slightly smaller than 2, which can be understood through two observations. First, the critical inverse temperatures in Eq.~\eqref{eq:transition-beta} indicate when $\pu>\pd$. However, in order to reach $\Ea$ from $\Eth$, the system needs to go through a large number of steps (i.e. growing with $N$) with ascending energy. If $\pu$ is only slightly larger than $\pd$ (which is what happens at $\beta$ slightly lower than 2), this is not enough to accept a sufficiently large number of steps to go all the way up to energies of order 1.
Second, this transition from barrier- to entropy-driven activation has features of a first-order phase transition. We see this from Fig.~\ref{fig:densities}, where we plot the distribution $P(\frac{\Er}{|\Eth|})$ at three different temperatures. At low $\beta$ it is peaked around 0, while at high $\beta$ it is peaked around $-1,$ as expected. Around the transition, instead, we see a two-peak structure that is characteristic of first-order phase transitions. We expect, then that the curves in Fig.~\ref{fig:ridges-beta}--top are subject to hysteresis. In that case, our measurement protocol, which starts measuring transitions when the system crosses $\Eth$, positions us in the barrier-driven metastable branch. Since we are measuring the highest-reached energy during a transition, it is non-trivial to position oneself in the entropic metastable branch.
We will extensively characterize these issues in future work~\cite{carbone:23}, by focusing on alternative observables such as the number of visited configurations during a transition.
Another potentially informative observable is the trapping time distribution. In Ref.~\onlinecite{baityjesi:18} the distribution of times spent below $\Eth$ in the EREM was reported to be $\psi_\mathrm{basin}(\tau)\sim \tau^{-\alpha_e}$, with $\alpha_e={1+\bc/\beta}$, which is what comes out from the Trap Model, which represents purely energetic activation. However, in the entropic phase of the EREM, there is a slow non-trivial dynamics above $\Eth$, in which the system's energy does not decrease easily due to a scarcity of descending directions. We would expect, then, that if instead of considering the times spent below $\Eth$ we consider those spent below $\Ea$, the distribution would widen and match the one predicted for the Step Model, $\psi_\mathrm{basin}\sim\tau^{-\alpha_a}$, with $\alpha_a={3-\beta/\bc}$~\cite{bertin:03}. However, this is hard to measure numerically, because the two exponents are very close to each other when $\bc<\beta<2\bc$. For example, at $\beta=\frac43\bc$, where most results of Ref.~\onlinecite{baityjesi:18} come from, we have $\alpha_e=1.75$ and $\alpha_a=1.6\overline{6}$. The maximum obtainable difference between the exponents is lower than 0.5, and it requires $\beta$ close to 2, where the entropic phase becomes spurious.

\paragraph*{Conclusions.}
We showed that two different kinds of activated dynamics, entropy- and barrier-driven, can coexist, though typically one dominates over the other. Barrier-driven activation corresponds to the typical picture of basin hopping, whereas entropy-driven activation is not driven by barrier heights, but rather by the scarcity of convenient directions. By studying two different forms of the density of states $\rho(E)$, we showed that, while at sufficiently low temperatures energetic activation always exists in landscapes with multiple minima, the existence of a higher-$T$ phase where entropy-driven activation dominates depends on the shape of $\rho(E)$. 
In this entropic activation phase, the attractor energy $\Ea$, towards which the system is regularly driven, is higher than the threshold energy $\Eth$. This means that in this phase the height of the barriers loses relevance.
As a consequence, the study of the transition paths of glasses through zero-temperature calculations, which is a technically daunting task~\cite{ros:19,ros:19b,ros:20,ros:21}, is potentially not informative for the dynamics at temperatures around the glass transition. The calculation of the attractor energy, with its comparison to the threshold, is a simpler calculation with the potential of unlocking the true activated nature of the dynamics. 

As a matter of fact, entropic activation can very elegantly explain an apparently puzzling result recently found by T. Rizzo~\cite{rizzo:21}, who calculated, in the spherical $p$-spin model, the path from one equilibrium low-temperature configuration to another, and found that the maximum energy reached during this transition is considerably higher than $\Eth$. Within the framework described here, we can hypothesize that the system is being pushed towards a higher energy $\Ea$ by the scarcity of paths with energy close to $\Eth$.
This is also consistent with recent numerical observations by Stariolo and Cugliandolo in the discrete $p$-spin model, that the trapping time distributions seem to follow the Step-model predictions better than those of the Trap Model~\cite{stariolo:19,stariolo:20}. In particular, they define the traps dynamically by taking, instead of the position of the saddle, the highest point reached during the dynamics. Therefore, they are calculating the traps through the attractor instead of the threshold energy, which explains the observed Step-like behavior. Furthermore, they find that the energy of the ridge is larger than $\Eth$, which is an indication that the lowest-energy path is not used, just as we find here.

Also in 3D systems, such a Lennard-Jones mixtures, although activated dynamics takes place between $\To$ and $\Td$~\cite{doliwa:03,doliwa:03c,doliwa:03d}, it was recently shown that it is not dominantly of an barrier-driven kind, since, for example, the system is moving at energies significantly higher than the ridges separating metabasins~\cite{baityjesi:21}: the height of the barriers separating basins does not play a crucial role in this regime. This is also what is pointed out in the dynamical facilitation picture, which shows that a strong glass-like slowdown can also appear in a trivial landscape, with the dynamics being slowed down by dynamical constraints~\cite{chandler:10,speck:19}. But entropic barriers \emph{are} dynamical constraints, since they effectively prevent the system from moving along directions which would be energetically favored. In other words, the entropic slowdown picture we find, which can also be present in trivial landscapes~\cite{bertin:03,bertin:10,bertin:12,cammarota:15,carbone:20}, is describing kinetic constraints. However, here, kinetic constraints are not the only slow dynamics mechanism, but they act and synergize with energetic activation; and both compete with the fast mechanism of diffusion towards lower energies.

Since the dynamical constraints appear at the onset temperature $\To$, we can assume that this corresponds to $1/\bc$ in the EREM. In the EREM, entropic activation stays dominant down to temperature $1/(2\bc)$, so an educated guess would be that the transition from entropic to barrier-driven occurs at the dynamical temperature $\Td$, given also past observations that a picture based on purely barrier-driven activation applies to supercooled liquids below $\Td$~\cite{schroder:00}. This also means that we do not expect entropic activation to be the dominant mechanism in glasses that do not have $\To>\Td$, such as the spherical $p$-spin model~\cite{cavagna:09}, or mean-field hard spheres~\cite{charbonneau:14}. On another hand, we expect that the entropic phase is present in the mixed $p$-spin model, where $\To>\Td$~\cite{folena:20,folena:21}.

In simple models such as the EREM or the Funnel model, correlations between neighboring configurations do not influence the presence of an entropically activated phase~\cite{carbone:20}: the phase is present both for local and global phase space dynamics. However, this is not necessarily true for more complex models such as the (pure and mixed) $p$-spin. A starting point which would allow to study correlated energy levels without changing the overall $\rho(E)$ would be the Correlated REM~\cite{baityjesi:18c} and the Number Partitioning Problem~\cite{junier:04}. A simple way to do this while still maintaining the exponential density of states is to study the Number Partitioning Problem~\cite{junier:04} or the Correlated REM~\cite{baityjesi:18c}. These two models have different kinds of correlations, since in the first, the traps are anti-correlated with their neighbors, while in the latter the basins are smooth.

Another question is attempting to relate the described mechanisms to real-space dynamics. By definition, barrier-driven activation requires a temporary increase of the energy. For example, this happens when a soft particle is squished between two other particles when passing from one side to the other. On the other side, a completely barrier-less transition would require many particles needing to cooperate in a collective movement. This would be the completely entropy-driven event, which would be very slow, as barrier hopping also is, but for a different reason. Since flat paths are not present in the EREM, the entropic effects that we observe are a compromise between the two, and would correspond to a collective rearrangement where the energy is slightly increased.

We close with a remark on the nature of activation in different kinds of models. 
Since all transitions in hard spheres consist of barrier-less collective rearrangements, do hard spheres even have barrier-hopping activation? Intuitively, this could pose a serious distinction between hard and soft models, given the qualitatively different kinds of activation that they can potentially assume.
Furthermore, a typical intuition of why mean-field results do not apply in low dimensions is that the energy barriers diverge with the system size. But if there are no energy barriers in hard spheres, are we sure that we are attributing the beyond-mean-field effects to the right mechanism? It was suggested in several occasions that the low-D correspondence of a mean-field model can be non-trivial~\cite{kirkpatrick:89,baityjesi:19b,folena:20}. 
Our analysis suggests that we should think of activation less as a hopping of energy barriers, and more as a search for convenient directions, which require a collective cooperative behavior that is hard to obtain by randomly moving particles. We should therefore regard activation as a process beyond MF, not because the barriers are diverging, but rather because it involves processes which take place in time scales $t\gg N$ (usually $t\sim e^N$). Thus, both entropic \emph{and} barrier-driven processes.
In this sense, there is a crucial difference between soft models (colloids, spin glasses, $\ldots$), where the underlying landscape is not flat, so at low enough temperature activation will eventually become barrier-driven, and hard models (such as hard spheres), where entropy-driven activation can persist all the way to $T=0.$

\section*{Acknowledgements}
We thank A. Altieri, V. Astuti, F. Ricci-Tersenghi, T. Rizzo, P. Sollich and D. Tapias for interesting conversations.
This work was funded by
the Simons Foundation for the collaboration Cracking
the Glass Problem (No. 454951 to D.R. Reichman).
M.~R.~C. acknowledges the following support: This material is based upon work supported by the U.S. Department of Energy, Office of Science, Office of Advanced Scientific Computing Research, Department of Energy Computational Science Graduate Fellowship under Award Number DE-FG02-97ER25308. 

\emph{Disclaimer}.--- This report was prepared as an account of work sponsored by an agency of the United States Government. Neither the United States Government nor any agency thereof, nor any of their employees, makes any warranty, express or implied, or assumes any legal liability or responsibility for the accuracy, completeness, or usefulness of any information, apparatus, product, or process disclosed, or represents that its use would not infringe privately owned rights. Reference herein to any specific commercial product, process, or service by trade name, trademark, manufacturer, or otherwise does not necessarily constitute or imply its endorsement, recommendation, or favoring by the United States Government or any agency thereof. The views and opinions of authors expressed herein do not necessarily state or reflect those of the United States Government or any agency thereof.

\appendix

\section{Numerical details}\label{app:numerical}
Our simulations are performed with the Metropolis Monte Carlo algorithm~\cite{metropolis:49}.
Depending on the value of $\beta$, we use two equivalent implementations, which we call ``standard" and ``Gillespie". The standard dynamics is a textbook Monte Carlo simulation with Metropolis acceptance criteria (see e.g. Ref.~\cite{krauth:06} for a detailed description). The run time of the standard procedure is tightly bound, with $t_\mathrm{max}$ time steps. However, at large $\beta,$ the rejection rate of standard dynamics is high and it may take many steps for a new configuration to be accepted. 
We therefore use the Gillespie procedure (which is formally equivalent) is much more efficient. 
The Gillespie algorithm is a rejectionless method, that computes the time that the system spends in a given configuration, and transitions without rejection to one of the neighbors, according to how probable it is to transition to each neighbor.
For a more detailed and didactic explanation of the Gillespie algorithm we refer the reader to Ref.~\cite{margiotta:18}. 
While the Gillespie method this is efficient at large $\beta,$ it is extremely inefficient at small $\beta.$ Thus, we set a cutoff of $\beta/\beta_\mathrm{c}=2.3,$ such that when $\beta/\beta_\mathrm{c} < 2.3,$ dynamics are run using the standard procedure, and Gillespie otherwise. We also use $t_\mathrm{max} = 10^7$ for all calculations.


\section{Memory and Memory-1 dynamics} \label{app:memory1}
In REM and EREM, the energy of each configuration is a fixed random variable. This means that, for $N$ spins, there are $2^N$ energies sampled from $\rho(E),$ each of which is permanently paired to some configuration. Therefore, in order to perform a long simulation in these models, we need to store the energy of all the $2^N$ states, to ensure that if a configuration is visited twice its energy has not changed. Storing $2^N$ double precision floating point numbers is expensive in terms of memory, and limits the largest system sizes that we can simulate. This is why, in order to simulate $N\geq30$, instead of storing the all the $2^N$ energies, we only stored the last visited one, and sampled anew the remaining $N-1$ neighbors. We call this dynamics \emph{memory-1}, in contrast with the \emph{memory} dynamics which stores all the energies throughout the whole simulation.

This simplification neglects loops in the dynamics, which for large $N$ are arguably rare, and does not allow the system to directly return to configurations visited more than one step earlier. The latter can be seen as an advantage, since we want to wash back and forth motion out of the dynamics we are measuring~\cite{doliwa:03c,baityjesi:18}.
Since the REM and EREM dynamics is a renewal process~\cite{baityjesi:18}, we can expect that anyhow after some time the previously visited phase space should be forgotten.

\begin{figure*} 
    \centering
    \includegraphics[width=\textwidth]{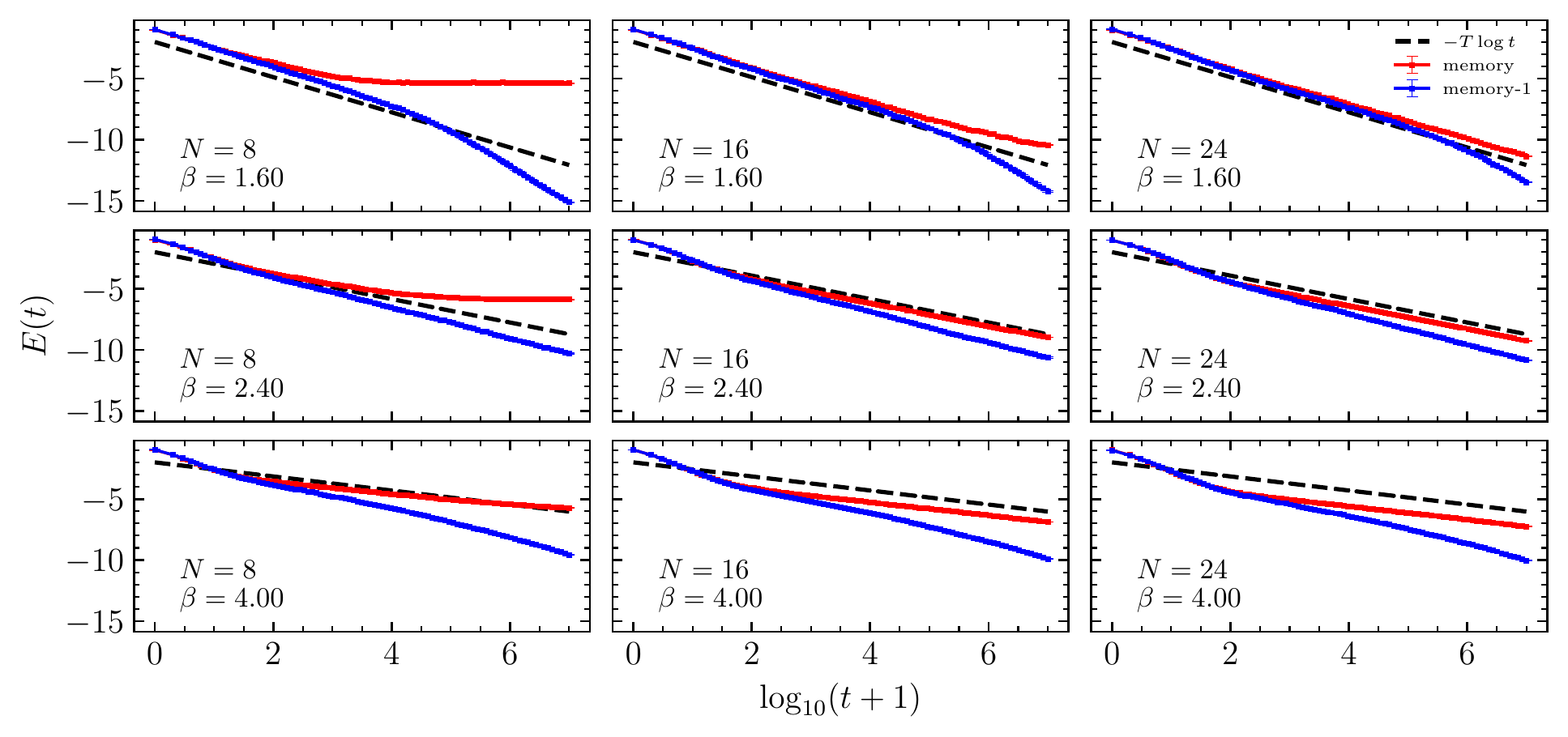}
    \caption{Energy as a function of time for memory and memory-1 dynamics. The dashed curve is the slope $-T\ln(t)$ that one would expect in the infinite-size limit.}
    \label{fig:memory1-Et}
\end{figure*}

An additional difference between memory and memory-1 dynamics is that the latter does not suffer from finite size effects descending from the phase space being of limited size: with memory dynamics there exists a lowest-reachable energy, while with memory-1 it is always possible to reach a lower energy. In other words, memory-1 dynamics suffer less from finite-size effects than the exact dynamics, and in any case this kind of effects does not affect the calculation of $\Er$.
In Fig.~\ref{fig:memory1-Et} we show the comparison between the two dynamics for varying $\beta$ and $N$.
At all temperatures, the difference between the two dynamics decreases as $N$ grows. Both dynamics present finite-size effects, which decrease as the system becomes larger.

\bibliographystyle{apsrev4-1}
\bibliography{marco}

\begin{thebibliography}{52}%
\makeatletter
\providecommand \@ifxundefined [1]{%
 \@ifx{#1\undefined}
}%
\providecommand \@ifnum [1]{%
 \ifnum #1\expandafter \@firstoftwo
 \else \expandafter \@secondoftwo
 \fi
}%
\providecommand \@ifx [1]{%
 \ifx #1\expandafter \@firstoftwo
 \else \expandafter \@secondoftwo
 \fi
}%
\providecommand \natexlab [1]{#1}%
\providecommand \enquote  [1]{``#1''}%
\providecommand \bibnamefont  [1]{#1}%
\providecommand \bibfnamefont [1]{#1}%
\providecommand \citenamefont [1]{#1}%
\providecommand \href@noop [0]{\@secondoftwo}%
\providecommand \href [0]{\begingroup \@sanitize@url \@href}%
\providecommand \@href[1]{\@@startlink{#1}\@@href}%
\providecommand \@@href[1]{\endgroup#1\@@endlink}%
\providecommand \@sanitize@url [0]{\catcode `\\12\catcode `\$12\catcode
  `\&12\catcode `\#12\catcode `\^12\catcode `\_12\catcode `\%12\relax}%
\providecommand \@@startlink[1]{}%
\providecommand \@@endlink[0]{}%
\providecommand \url  [0]{\begingroup\@sanitize@url \@url }%
\providecommand \@url [1]{\endgroup\@href {#1}{\urlprefix }}%
\providecommand \urlprefix  [0]{URL }%
\providecommand \Eprint [0]{\href }%
\providecommand \doibase [0]{http://dx.doi.org/}%
\providecommand \selectlanguage [0]{\@gobble}%
\providecommand \bibinfo  [0]{\@secondoftwo}%
\providecommand \bibfield  [0]{\@secondoftwo}%
\providecommand \translation [1]{[#1]}%
\providecommand \BibitemOpen [0]{}%
\providecommand \bibitemStop [0]{}%
\providecommand \bibitemNoStop [0]{.\EOS\space}%
\providecommand \EOS [0]{\spacefactor3000\relax}%
\providecommand \BibitemShut  [1]{\csname bibitem#1\endcsname}%
\let\auto@bib@innerbib\@empty
\bibitem [{\citenamefont {Charbonneau}\ \emph {et~al.}(2014)\citenamefont
  {Charbonneau}, \citenamefont {Kurchan}, \citenamefont {Parisi}, \citenamefont
  {Urbani},\ and\ \citenamefont {Zamponi}}]{charbonneau:14}%
  \BibitemOpen
  \bibfield  {author} {\bibinfo {author} {\bibfnamefont {P.}~\bibnamefont
  {Charbonneau}}, \bibinfo {author} {\bibfnamefont {J.}~\bibnamefont
  {Kurchan}}, \bibinfo {author} {\bibfnamefont {G.}~\bibnamefont {Parisi}},
  \bibinfo {author} {\bibfnamefont {P.}~\bibnamefont {Urbani}}, \ and\ \bibinfo
  {author} {\bibfnamefont {F.}~\bibnamefont {Zamponi}},\ }\href {\doibase
  10.1038/ncomms4725} {\bibfield  {journal} {\bibinfo  {journal} {Nature
  Communications}\ }\textbf {\bibinfo {volume} {5}},\ \bibinfo {pages} {3725}
  (\bibinfo {year} {2014})},\ \Eprint {http://arxiv.org/abs/arXiv:1404.6809}
  {arXiv:1404.6809} \BibitemShut {NoStop}%
\bibitem [{\citenamefont {Maimbourg}\ \emph {et~al.}(2016)\citenamefont
  {Maimbourg}, \citenamefont {Kurchan},\ and\ \citenamefont
  {Zamponi}}]{maimbourg:16}%
  \BibitemOpen
  \bibfield  {author} {\bibinfo {author} {\bibfnamefont {T.}~\bibnamefont
  {Maimbourg}}, \bibinfo {author} {\bibfnamefont {J.}~\bibnamefont {Kurchan}},
  \ and\ \bibinfo {author} {\bibfnamefont {F.}~\bibnamefont {Zamponi}},\ }\href
  {\doibase https://doi.org/10.1103/PhysRevLett.116.015902} {\bibfield
  {journal} {\bibinfo  {journal} {Physical review letters}\ }\textbf {\bibinfo
  {volume} {116}},\ \bibinfo {pages} {015902} (\bibinfo {year}
  {2016})}\BibitemShut {NoStop}%
\bibitem [{\citenamefont {Charbonneau}\ \emph {et~al.}(2017)\citenamefont
  {Charbonneau}, \citenamefont {Kurchan}, \citenamefont {Parisi}, \citenamefont
  {Urbani},\ and\ \citenamefont {Zamponi}}]{charbonneau:17}%
  \BibitemOpen
  \bibfield  {author} {\bibinfo {author} {\bibfnamefont {P.}~\bibnamefont
  {Charbonneau}}, \bibinfo {author} {\bibfnamefont {J.}~\bibnamefont
  {Kurchan}}, \bibinfo {author} {\bibfnamefont {G.}~\bibnamefont {Parisi}},
  \bibinfo {author} {\bibfnamefont {P.}~\bibnamefont {Urbani}}, \ and\ \bibinfo
  {author} {\bibfnamefont {F.}~\bibnamefont {Zamponi}},\ }\href {\doibase
  https://doi.org/10.1146/annurev-conmatphys-031016-025334} {\bibfield
  {journal} {\bibinfo  {journal} {Annual Review of Condensed Matter Physics}\
  }\textbf {\bibinfo {volume} {8}},\ \bibinfo {pages} {265} (\bibinfo {year}
  {2017})}\BibitemShut {NoStop}%
\bibitem [{\citenamefont {Doliwa}\ and\ \citenamefont
  {Heuer}(2003{\natexlab{a}})}]{doliwa:03c}%
  \BibitemOpen
  \bibfield  {author} {\bibinfo {author} {\bibfnamefont {B.}~\bibnamefont
  {Doliwa}}\ and\ \bibinfo {author} {\bibfnamefont {A.}~\bibnamefont {Heuer}},\
  }\href@noop {} {\bibfield  {journal} {\bibinfo  {journal} {Physical Review
  E}\ }\textbf {\bibinfo {volume} {67}},\ \bibinfo {pages} {031506} (\bibinfo
  {year} {2003}{\natexlab{a}})},\ \Eprint
  {http://arxiv.org/abs/arxiv:cond-mat/0209139} {arxiv:cond-mat/0209139}
  \BibitemShut {NoStop}%
\bibitem [{\citenamefont {Arceri}\ \emph {et~al.}(2020)\citenamefont {Arceri},
  \citenamefont {Landes}, \citenamefont {Berthier},\ and\ \citenamefont
  {Biroli}}]{arceri:20}%
  \BibitemOpen
  \bibfield  {author} {\bibinfo {author} {\bibfnamefont {F.}~\bibnamefont
  {Arceri}}, \bibinfo {author} {\bibfnamefont {F.~P.}\ \bibnamefont {Landes}},
  \bibinfo {author} {\bibfnamefont {L.}~\bibnamefont {Berthier}}, \ and\
  \bibinfo {author} {\bibfnamefont {G.}~\bibnamefont {Biroli}},\ }\href@noop {}
  {\enquote {\bibinfo {title} {Glasses and aging: a statistical mechanics
  perspective},}\ } (\bibinfo {year} {2020}),\ \Eprint
  {http://arxiv.org/abs/arXiv:2006.09725} {arXiv:2006.09725} \BibitemShut
  {NoStop}%
\bibitem [{\citenamefont {Crisanti}\ and\ \citenamefont
  {Ritort}(2000{\natexlab{a}})}]{crisanti:00}%
  \BibitemOpen
  \bibfield  {author} {\bibinfo {author} {\bibfnamefont {A.}~\bibnamefont
  {Crisanti}}\ and\ \bibinfo {author} {\bibfnamefont {F.}~\bibnamefont
  {Ritort}},\ }\href {\doibase https://doi.org/10.1209/epl/i2000-00486-2}
  {\bibfield  {journal} {\bibinfo  {journal} {EPL (Europhysics Letters)}\
  }\textbf {\bibinfo {volume} {52}},\ \bibinfo {pages} {640} (\bibinfo {year}
  {2000}{\natexlab{a}})}\BibitemShut {NoStop}%
\bibitem [{\citenamefont {Crisanti}\ and\ \citenamefont
  {Ritort}(2000{\natexlab{b}})}]{crisanti:00b}%
  \BibitemOpen
  \bibfield  {author} {\bibinfo {author} {\bibfnamefont {A.}~\bibnamefont
  {Crisanti}}\ and\ \bibinfo {author} {\bibfnamefont {F.}~\bibnamefont
  {Ritort}},\ }\href {http://stacks.iop.org/0295-5075/51/i=2/a=147} {\bibfield
  {journal} {\bibinfo  {journal} {EPL (Europhysics Letters)}\ }\textbf
  {\bibinfo {volume} {51}},\ \bibinfo {pages} {147} (\bibinfo {year}
  {2000}{\natexlab{b}})}\BibitemShut {NoStop}%
\bibitem [{\citenamefont {Derrida}(1980)}]{derrida:80}%
  \BibitemOpen
  \bibfield  {author} {\bibinfo {author} {\bibfnamefont {B.}~\bibnamefont
  {Derrida}},\ }\href {\doibase 10.1103/PhysRevLett.45.79} {\bibfield
  {journal} {\bibinfo  {journal} {Phys. Rev. Lett.}\ }\textbf {\bibinfo
  {volume} {45}},\ \bibinfo {pages} {79} (\bibinfo {year} {1980})}\BibitemShut
  {NoStop}%
\bibitem [{\citenamefont {Dyre}(1987)}]{dyre:87}%
  \BibitemOpen
  \bibfield  {author} {\bibinfo {author} {\bibfnamefont {J.~C.}\ \bibnamefont
  {Dyre}},\ }\href {\doibase 10.1103/PhysRevLett.58.792} {\bibfield  {journal}
  {\bibinfo  {journal} {Phys. Rev. Lett.}\ }\textbf {\bibinfo {volume} {58}},\
  \bibinfo {pages} {792} (\bibinfo {year} {1987})}\BibitemShut {NoStop}%
\bibitem [{\citenamefont {Bouchaud}(1992)}]{bouchaud:92}%
  \BibitemOpen
  \bibfield  {author} {\bibinfo {author} {\bibfnamefont {J.}~\bibnamefont
  {Bouchaud}},\ }\href {\doibase 10.1051/jp1:1992238} {\bibfield  {journal}
  {\bibinfo  {journal} {J. Phys. I France}\ }\textbf {\bibinfo {volume} {2}},\
  \bibinfo {pages} {1705} (\bibinfo {year} {1992})}\BibitemShut {NoStop}%
\bibitem [{\citenamefont {{J.-P. Bouchaud}}\ and\ \citenamefont {{D.S.
  Dean}}(1995)}]{bouchaud:95}%
  \BibitemOpen
  \bibfield  {author} {\bibinfo {author} {\bibnamefont {{J.-P. Bouchaud}}}\
  and\ \bibinfo {author} {\bibnamefont {{D.S. Dean}}},\ }\href {\doibase
  10.1051/jp1:1995127} {\bibfield  {journal} {\bibinfo  {journal} {J. Phys. I
  France}\ }\textbf {\bibinfo {volume} {5}},\ \bibinfo {pages} {265} (\bibinfo
  {year} {1995})},\ \Eprint {http://arxiv.org/abs/arXiv:cond-mat/9410022}
  {arXiv:cond-mat/9410022} \BibitemShut {NoStop}%
\bibitem [{\citenamefont {Ben-Arous}\ \emph {et~al.}(2002)\citenamefont
  {Ben-Arous}, \citenamefont {Bovier},\ and\ \citenamefont
  {Gayrard}}]{benarous:02}%
  \BibitemOpen
  \bibfield  {author} {\bibinfo {author} {\bibfnamefont {G.}~\bibnamefont
  {Ben-Arous}}, \bibinfo {author} {\bibfnamefont {A.}~\bibnamefont {Bovier}}, \
  and\ \bibinfo {author} {\bibfnamefont {V.}~\bibnamefont {Gayrard}},\ }\href
  {\doibase 10.1103/PhysRevLett.88.087201} {\bibfield  {journal} {\bibinfo
  {journal} {Phys. Rev. Lett.}\ }\textbf {\bibinfo {volume} {88}},\ \bibinfo
  {pages} {087201} (\bibinfo {year} {2002})}\BibitemShut {NoStop}%
\bibitem [{\citenamefont {Ben-Arous}\ \emph {et~al.}(2008)\citenamefont
  {Ben-Arous}, \citenamefont {Bovier},\ and\ \citenamefont
  {{\v{C}}ern{\`y}}}]{benarous:08}%
  \BibitemOpen
  \bibfield  {author} {\bibinfo {author} {\bibfnamefont {G.}~\bibnamefont
  {Ben-Arous}}, \bibinfo {author} {\bibfnamefont {A.}~\bibnamefont {Bovier}}, \
  and\ \bibinfo {author} {\bibfnamefont {J.}~\bibnamefont {{\v{C}}ern{\`y}}},\
  }\href {\doibase 10.1007/s00220-008-0565-7} {\bibfield  {journal} {\bibinfo
  {journal} {Communications in Mathematical Physics}\ }\textbf {\bibinfo
  {volume} {282}},\ \bibinfo {pages} {663} (\bibinfo {year}
  {2008})}\BibitemShut {NoStop}%
\bibitem [{\citenamefont {Gayrard}(2016)}]{gayrard:16}%
  \BibitemOpen
  \bibfield  {author} {\bibinfo {author} {\bibfnamefont {V.}~\bibnamefont
  {Gayrard}},\ }\href {\doibase 10.1007/s00023-015-0442-9} {\bibfield
  {journal} {\bibinfo  {journal} {Annales Henri Poincar{\'e}}\ }\textbf
  {\bibinfo {volume} {17}},\ \bibinfo {pages} {537} (\bibinfo {year}
  {2016})}\BibitemShut {NoStop}%
\bibitem [{\citenamefont {Gayrard}\ and\ \citenamefont
  {G{\"u}n}(2016)}]{gayrard:16b}%
  \BibitemOpen
  \bibfield  {author} {\bibinfo {author} {\bibfnamefont {V.}~\bibnamefont
  {Gayrard}}\ and\ \bibinfo {author} {\bibfnamefont {O.}~\bibnamefont
  {G{\"u}n}},\ }\href@noop {} {\bibfield  {journal} {\bibinfo  {journal}
  {Markov Processes and Related Fields}\ }\textbf {\bibinfo {volume} {22}},\
  \bibinfo {pages} {165} (\bibinfo {year} {2016})}\BibitemShut {NoStop}%
\bibitem [{\citenamefont {{\v{C}}ern{\'y}}\ and\ \citenamefont
  {Wassmer}(2017)}]{cerny:17}%
  \BibitemOpen
  \bibfield  {author} {\bibinfo {author} {\bibfnamefont {J.}~\bibnamefont
  {{\v{C}}ern{\'y}}}\ and\ \bibinfo {author} {\bibfnamefont {T.}~\bibnamefont
  {Wassmer}},\ }\href {\doibase 10.1007/s00440-015-0681-1} {\bibfield
  {journal} {\bibinfo  {journal} {Probab. Theory Relat. Fields}\ }\textbf
  {\bibinfo {volume} {167}},\ \bibinfo {pages} {253} (\bibinfo {year}
  {2017})}\BibitemShut {NoStop}%
\bibitem [{\citenamefont {Gayrard}(2018)}]{gayrard:18}%
  \BibitemOpen
  \bibfield  {author} {\bibinfo {author} {\bibfnamefont {V.}~\bibnamefont
  {Gayrard}},\ }\href {\doibase 10.1007/s00440-018-0873-6} {\bibfield
  {journal} {\bibinfo  {journal} {Probab. Theory Relat. Fields}\ } (\bibinfo
  {year} {2018}),\ 10.1007/s00440-018-0873-6},\ \Eprint
  {http://arxiv.org/abs/arXiv:1602.06081} {arXiv:1602.06081} \BibitemShut
  {NoStop}%
\bibitem [{\citenamefont {Gayrard}\ and\ \citenamefont
  {Hartung}(2019)}]{gayrard:19}%
  \BibitemOpen
  \bibfield  {author} {\bibinfo {author} {\bibfnamefont {V.}~\bibnamefont
  {Gayrard}}\ and\ \bibinfo {author} {\bibfnamefont {L.}~\bibnamefont
  {Hartung}},\ }in\ \href@noop {} {\emph {\bibinfo {booktitle} {Statistical
  Mechanics of Classical and Disordered Systems}}},\ \bibinfo {editor} {edited
  by\ \bibinfo {editor} {\bibfnamefont {V.}~\bibnamefont {Gayrard}}, \bibinfo
  {editor} {\bibfnamefont {L.-P.}\ \bibnamefont {Arguin}}, \bibinfo {editor}
  {\bibfnamefont {N.}~\bibnamefont {Kistler}}, \ and\ \bibinfo {editor}
  {\bibfnamefont {I.}~\bibnamefont {Kourkova}}}\ (\bibinfo  {publisher}
  {Springer International Publishing},\ \bibinfo {address} {Cham},\ \bibinfo
  {year} {2019})\ pp.\ \bibinfo {pages} {111--170}\BibitemShut {NoStop}%
\bibitem [{\citenamefont {Baity-Jesi}\ \emph
  {et~al.}(2018{\natexlab{a}})\citenamefont {Baity-Jesi}, \citenamefont
  {Biroli},\ and\ \citenamefont {Cammarota}}]{baityjesi:18}%
  \BibitemOpen
  \bibfield  {author} {\bibinfo {author} {\bibfnamefont {M.}~\bibnamefont
  {Baity-Jesi}}, \bibinfo {author} {\bibfnamefont {G.}~\bibnamefont {Biroli}},
  \ and\ \bibinfo {author} {\bibfnamefont {C.}~\bibnamefont {Cammarota}},\
  }\href {http://stacks.iop.org/1742-5468/2018/i=1/a=013301} {\bibfield
  {journal} {\bibinfo  {journal} {J. Stat. Mech.: Theory Exp}\ ,\ \bibinfo
  {pages} {013301}} (\bibinfo {year} {2018}{\natexlab{a}})},\ \Eprint
  {http://arxiv.org/abs/arXiv:1708.03268} {arXiv:1708.03268} \BibitemShut
  {NoStop}%
\bibitem [{\citenamefont {Baity-Jesi}\ \emph
  {et~al.}(2018{\natexlab{b}})\citenamefont {Baity-Jesi}, \citenamefont
  {Achard-de Lustrac},\ and\ \citenamefont {Biroli}}]{baityjesi:18c}%
  \BibitemOpen
  \bibfield  {author} {\bibinfo {author} {\bibfnamefont {M.}~\bibnamefont
  {Baity-Jesi}}, \bibinfo {author} {\bibfnamefont {A.}~\bibnamefont {Achard-de
  Lustrac}}, \ and\ \bibinfo {author} {\bibfnamefont {G.}~\bibnamefont
  {Biroli}},\ }\href@noop {} {\bibfield  {journal} {\bibinfo  {journal} {Phys.
  Rev. E}\ }\textbf {\bibinfo {volume} {98}},\ \bibinfo {pages} {012133}
  (\bibinfo {year} {2018}{\natexlab{b}})},\ \Eprint
  {http://arxiv.org/abs/arXiv:1805.04581} {arXiv:1805.04581} \BibitemShut
  {NoStop}%
\bibitem [{\citenamefont {Ros}\ \emph {et~al.}(2019{\natexlab{a}})\citenamefont
  {Ros}, \citenamefont {Arous}, \citenamefont {Biroli},\ and\ \citenamefont
  {Cammarota}}]{ros:19}%
  \BibitemOpen
  \bibfield  {author} {\bibinfo {author} {\bibfnamefont {V.}~\bibnamefont
  {Ros}}, \bibinfo {author} {\bibfnamefont {G.~B.}\ \bibnamefont {Arous}},
  \bibinfo {author} {\bibfnamefont {G.}~\bibnamefont {Biroli}}, \ and\ \bibinfo
  {author} {\bibfnamefont {C.}~\bibnamefont {Cammarota}},\ }\href@noop {}
  {\bibfield  {journal} {\bibinfo  {journal} {Physical Review X}\ }\textbf
  {\bibinfo {volume} {9}},\ \bibinfo {pages} {011003} (\bibinfo {year}
  {2019}{\natexlab{a}})},\ \Eprint {http://arxiv.org/abs/arXiv:1804.02686}
  {arXiv:1804.02686} \BibitemShut {NoStop}%
\bibitem [{\citenamefont {Ros}\ \emph {et~al.}(2019{\natexlab{b}})\citenamefont
  {Ros}, \citenamefont {Biroli},\ and\ \citenamefont {Cammarota}}]{ros:19b}%
  \BibitemOpen
  \bibfield  {author} {\bibinfo {author} {\bibfnamefont {V.}~\bibnamefont
  {Ros}}, \bibinfo {author} {\bibfnamefont {G.}~\bibnamefont {Biroli}}, \ and\
  \bibinfo {author} {\bibfnamefont {C.}~\bibnamefont {Cammarota}},\ }\href@noop
  {} {\bibfield  {journal} {\bibinfo  {journal} {EPL (Europhysics Letters)}\
  }\textbf {\bibinfo {volume} {126}},\ \bibinfo {pages} {20003} (\bibinfo
  {year} {2019}{\natexlab{b}})}\BibitemShut {NoStop}%
\bibitem [{\citenamefont {Ros}(2020)}]{ros:20}%
  \BibitemOpen
  \bibfield  {author} {\bibinfo {author} {\bibfnamefont {V.}~\bibnamefont
  {Ros}},\ }\href@noop {} {\bibfield  {journal} {\bibinfo  {journal} {Journal
  of Physics A: Mathematical and Theoretical}\ }\textbf {\bibinfo {volume}
  {53}},\ \bibinfo {pages} {125002} (\bibinfo {year} {2020})}\BibitemShut
  {NoStop}%
\bibitem [{\citenamefont {Ros}\ \emph {et~al.}(2021)\citenamefont {Ros},
  \citenamefont {Biroli},\ and\ \citenamefont {Cammarota}}]{ros:21}%
  \BibitemOpen
  \bibfield  {author} {\bibinfo {author} {\bibfnamefont {V.}~\bibnamefont
  {Ros}}, \bibinfo {author} {\bibfnamefont {G.}~\bibnamefont {Biroli}}, \ and\
  \bibinfo {author} {\bibfnamefont {C.}~\bibnamefont {Cammarota}},\ }\href@noop
  {} {\bibfield  {journal} {\bibinfo  {journal} {SciPost Physics}\ }\textbf
  {\bibinfo {volume} {10}},\ \bibinfo {pages} {002} (\bibinfo {year}
  {2021})}\BibitemShut {NoStop}%
\bibitem [{\citenamefont {Rizzo}(2021)}]{rizzo:21}%
  \BibitemOpen
  \bibfield  {author} {\bibinfo {author} {\bibfnamefont {T.}~\bibnamefont
  {Rizzo}},\ }\href {\doibase 10.1103/PhysRevB.104.094203} {\bibfield
  {journal} {\bibinfo  {journal} {Physical Review B}\ }\textbf {\bibinfo
  {volume} {104}},\ \bibinfo {pages} {094203} (\bibinfo {year} {2021})},\
  \Eprint {http://arxiv.org/abs/arXiv:2012.09556} {arXiv:2012.09556}
  \BibitemShut {NoStop}%
\bibitem [{\citenamefont {Cammarota}\ and\ \citenamefont
  {Marinari}(2015)}]{cammarota:15}%
  \BibitemOpen
  \bibfield  {author} {\bibinfo {author} {\bibfnamefont {C.}~\bibnamefont
  {Cammarota}}\ and\ \bibinfo {author} {\bibfnamefont {E.}~\bibnamefont
  {Marinari}},\ }\href {\doibase https://doi.org/10.1103/PhysRevE.92.010301}
  {\bibfield  {journal} {\bibinfo  {journal} {Phys. Rev. E}\ }\textbf {\bibinfo
  {volume} {92}},\ \bibinfo {pages} {010301(R)} (\bibinfo {year} {2015})},\
  \Eprint {http://arxiv.org/abs/arXiv:1410.2116} {arXiv:1410.2116} \BibitemShut
  {NoStop}%
\bibitem [{\citenamefont {Carbone}\ \emph {et~al.}(2020)\citenamefont
  {Carbone}, \citenamefont {Astuti},\ and\ \citenamefont
  {Baity-Jesi}}]{carbone:20}%
  \BibitemOpen
  \bibfield  {author} {\bibinfo {author} {\bibfnamefont {M.~R.}\ \bibnamefont
  {Carbone}}, \bibinfo {author} {\bibfnamefont {V.}~\bibnamefont {Astuti}}, \
  and\ \bibinfo {author} {\bibfnamefont {M.}~\bibnamefont {Baity-Jesi}},\
  }\href {\doibase 10.1103/PhysRevE.101.052304} {\bibfield  {journal} {\bibinfo
   {journal} {Phys. Rev. E}\ }\textbf {\bibinfo {volume} {101}},\ \bibinfo
  {pages} {052304} (\bibinfo {year} {2020})},\ \Eprint
  {http://arxiv.org/abs/arXiv:2001.02567} {arXiv:2001.02567} \BibitemShut
  {NoStop}%
\bibitem [{\citenamefont {Stariolo}\ and\ \citenamefont
  {Cugliandolo}(2019)}]{stariolo:19}%
  \BibitemOpen
  \bibfield  {author} {\bibinfo {author} {\bibfnamefont {D.~A.}\ \bibnamefont
  {Stariolo}}\ and\ \bibinfo {author} {\bibfnamefont {L.~F.}\ \bibnamefont
  {Cugliandolo}},\ }\href {\doibase 10.1209/0295-5075/127/16002} {\bibfield
  {journal} {\bibinfo  {journal} {{EPL} (Europhysics Letters)}\ }\textbf
  {\bibinfo {volume} {127}},\ \bibinfo {pages} {16002} (\bibinfo {year}
  {2019})},\ \Eprint {http://arxiv.org/abs/arXiv:1904.10731} {arXiv:1904.10731}
  \BibitemShut {NoStop}%
\bibitem [{\citenamefont {Stariolo}\ and\ \citenamefont
  {Cugliandolo}(2020)}]{stariolo:20}%
  \BibitemOpen
  \bibfield  {author} {\bibinfo {author} {\bibfnamefont {D.~A.}\ \bibnamefont
  {Stariolo}}\ and\ \bibinfo {author} {\bibfnamefont {L.~F.}\ \bibnamefont
  {Cugliandolo}},\ }\href {\doibase 10.1103/PhysRevE.102.022126} {\bibfield
  {journal} {\bibinfo  {journal} {Phys. Rev. E}\ }\textbf {\bibinfo {volume}
  {102}},\ \bibinfo {pages} {022126} (\bibinfo {year} {2020})},\ \Eprint
  {http://arxiv.org/abs/arXiv:2004.09410} {arXiv:2004.09410} \BibitemShut
  {NoStop}%
\bibitem [{\citenamefont {Tapias}\ \emph {et~al.}(2020)\citenamefont {Tapias},
  \citenamefont {Paprotzki},\ and\ \citenamefont {Sollich}}]{tapias:20}%
  \BibitemOpen
  \bibfield  {author} {\bibinfo {author} {\bibfnamefont {D.}~\bibnamefont
  {Tapias}}, \bibinfo {author} {\bibfnamefont {E.}~\bibnamefont {Paprotzki}}, \
  and\ \bibinfo {author} {\bibfnamefont {P.}~\bibnamefont {Sollich}},\ }\href
  {\doibase 10.1088/1742-5468/abaecf} {\bibfield  {journal} {\bibinfo
  {journal} {Journal of Statistical Mechanics: Theory and Experiment}\ }\textbf
  {\bibinfo {volume} {2020}},\ \bibinfo {pages} {093302} (\bibinfo {year}
  {2020})},\ \Eprint {http://arxiv.org/abs/arXiv:2005.04994} {arXiv:2005.04994}
  \BibitemShut {NoStop}%
\bibitem [{\citenamefont {Baity-Jesi}\ \emph {et~al.}(2021)\citenamefont
  {Baity-Jesi}, \citenamefont {Biroli},\ and\ \citenamefont
  {Reichman}}]{baityjesi:21}%
  \BibitemOpen
  \bibfield  {author} {\bibinfo {author} {\bibfnamefont {M.}~\bibnamefont
  {Baity-Jesi}}, \bibinfo {author} {\bibfnamefont {G.}~\bibnamefont {Biroli}},
  \ and\ \bibinfo {author} {\bibfnamefont {D.~R.}\ \bibnamefont {Reichman}},\
  }\href {\doibase 10.1140/epje/s10189-021-00077-y} {\bibfield  {journal}
  {\bibinfo  {journal} {The European Physical Journal E}\ }\textbf {\bibinfo
  {volume} {44}},\ \bibinfo {pages} {1} (\bibinfo {year} {2021})},\ \Eprint
  {http://arxiv.org/abs/arXiv:2103.07211} {arXiv:2103.07211} \BibitemShut
  {NoStop}%
\bibitem [{\citenamefont {Chandler}\ and\ \citenamefont
  {Garrahan}(2010)}]{chandler:10}%
  \BibitemOpen
  \bibfield  {author} {\bibinfo {author} {\bibfnamefont {D.}~\bibnamefont
  {Chandler}}\ and\ \bibinfo {author} {\bibfnamefont {J.~P.}\ \bibnamefont
  {Garrahan}},\ }\href {\doibase 10.1146/annurev.physchem.040808.090405}
  {\bibfield  {journal} {\bibinfo  {journal} {Annual review of physical
  chemistry}\ }\textbf {\bibinfo {volume} {61}},\ \bibinfo {pages} {191}
  (\bibinfo {year} {2010})}\BibitemShut {NoStop}%
\bibitem [{\citenamefont {Royall}\ \emph {et~al.}(2020)\citenamefont {Royall},
  \citenamefont {Turci},\ and\ \citenamefont {Speck}}]{royall:20}%
  \BibitemOpen
  \bibfield  {author} {\bibinfo {author} {\bibfnamefont {C.~P.}\ \bibnamefont
  {Royall}}, \bibinfo {author} {\bibfnamefont {F.}~\bibnamefont {Turci}}, \
  and\ \bibinfo {author} {\bibfnamefont {T.}~\bibnamefont {Speck}},\ }\href
  {\doibase 10.1063/5.0006998} {\bibfield  {journal} {\bibinfo  {journal} {The
  Journal of Chemical Physics}\ }\textbf {\bibinfo {volume} {153}},\ \bibinfo
  {pages} {090901} (\bibinfo {year} {2020})},\ \Eprint
  {http://arxiv.org/abs/arXiv:2003.03700} {arXiv:2003.03700} \BibitemShut
  {NoStop}%
\bibitem [{\citenamefont {Castellani}\ and\ \citenamefont
  {Cavagna}(2005)}]{castellani:05}%
  \BibitemOpen
  \bibfield  {author} {\bibinfo {author} {\bibfnamefont {T.}~\bibnamefont
  {Castellani}}\ and\ \bibinfo {author} {\bibfnamefont {A.}~\bibnamefont
  {Cavagna}},\ }\href {\doibase 10.1088/1742-5468/2005/05/P05012} {\bibfield
  {journal} {\bibinfo  {journal} {J. Stat. Mech.}\ }\textbf {\bibinfo {volume}
  {2005}},\ \bibinfo {pages} {P05012} (\bibinfo {year} {2005})}\BibitemShut
  {NoStop}%
\bibitem [{\citenamefont {Hartarsky}\ \emph {et~al.}(2019)\citenamefont
  {Hartarsky}, \citenamefont {Baity-Jesi}, \citenamefont {Ravasio},
  \citenamefont {Billoire},\ and\ \citenamefont {Biroli}}]{hartarsky:19}%
  \BibitemOpen
  \bibfield  {author} {\bibinfo {author} {\bibfnamefont {I.}~\bibnamefont
  {Hartarsky}}, \bibinfo {author} {\bibfnamefont {M.}~\bibnamefont
  {Baity-Jesi}}, \bibinfo {author} {\bibfnamefont {R.}~\bibnamefont {Ravasio}},
  \bibinfo {author} {\bibfnamefont {A.}~\bibnamefont {Billoire}}, \ and\
  \bibinfo {author} {\bibfnamefont {G.}~\bibnamefont {Biroli}},\ }\href
  {\doibase 10.1088/1742-5468/ab38bf} {\bibfield  {journal} {\bibinfo
  {journal} {J. Stat. Mech.: Theory Exp}\ }\textbf {\bibinfo {volume} {2019}},\
  \bibinfo {pages} {093302} (\bibinfo {year} {2019})},\ \Eprint
  {http://arxiv.org/abs/arXiv:1904.08024} {arXiv:1904.08024} \BibitemShut
  {NoStop}%
\bibitem [{\citenamefont {Barrat}\ and\ \citenamefont
  {M{\'e}zard}(1995)}]{barrat:95}%
  \BibitemOpen
  \bibfield  {author} {\bibinfo {author} {\bibfnamefont {A.}~\bibnamefont
  {Barrat}}\ and\ \bibinfo {author} {\bibfnamefont {M.}~\bibnamefont
  {M{\'e}zard}},\ }\href {\doibase 10.1051/jp1:1995174} {\bibfield  {journal}
  {\bibinfo  {journal} {Journal de Physique I}\ }\textbf {\bibinfo {volume}
  {5}},\ \bibinfo {pages} {941} (\bibinfo {year} {1995})},\ \Eprint
  {http://arxiv.org/abs/cond-mat/9504088} {cond-mat/9504088} \BibitemShut
  {NoStop}%
\bibitem [{\citenamefont {Bertin}(2003)}]{bertin:03}%
  \BibitemOpen
  \bibfield  {author} {\bibinfo {author} {\bibfnamefont {E.~M.}\ \bibnamefont
  {Bertin}},\ }\href {\doibase 10.1088/0305-4470/36/43/002} {\bibfield
  {journal} {\bibinfo  {journal} {J. Phys. A}\ }\textbf {\bibinfo {volume}
  {36}},\ \bibinfo {pages} {10683} (\bibinfo {year} {2003})},\ \Eprint
  {http://arxiv.org/abs/arXiv:cond-mat/0305538} {arXiv:cond-mat/0305538}
  \BibitemShut {NoStop}%
\bibitem [{\citenamefont {Doliwa}\ and\ \citenamefont
  {Heuer}(2003{\natexlab{b}})}]{doliwa:03}%
  \BibitemOpen
  \bibfield  {author} {\bibinfo {author} {\bibfnamefont {B.}~\bibnamefont
  {Doliwa}}\ and\ \bibinfo {author} {\bibfnamefont {A.}~\bibnamefont {Heuer}},\
  }\href {\doibase 10.1103/PhysRevLett.91.235501} {\bibfield  {journal}
  {\bibinfo  {journal} {Phys. Rev. Lett.}\ }\textbf {\bibinfo {volume} {91}},\
  \bibinfo {pages} {235501} (\bibinfo {year} {2003}{\natexlab{b}})}\BibitemShut
  {NoStop}%
\bibitem [{\citenamefont {Doliwa}\ and\ \citenamefont
  {Heuer}(2003{\natexlab{c}})}]{doliwa:03d}%
  \BibitemOpen
  \bibfield  {author} {\bibinfo {author} {\bibfnamefont {B.}~\bibnamefont
  {Doliwa}}\ and\ \bibinfo {author} {\bibfnamefont {A.}~\bibnamefont {Heuer}},\
  }\href@noop {} {\bibfield  {journal} {\bibinfo  {journal} {Physical Review
  E}\ }\textbf {\bibinfo {volume} {67}},\ \bibinfo {pages} {030501} (\bibinfo
  {year} {2003}{\natexlab{c}})}\BibitemShut {NoStop}%
\bibitem [{\citenamefont {Speck}(2019)}]{speck:19}%
  \BibitemOpen
  \bibfield  {author} {\bibinfo {author} {\bibfnamefont {T.}~\bibnamefont
  {Speck}},\ }\href@noop {} {\bibfield  {journal} {\bibinfo  {journal} {Journal
  of Statistical Mechanics: Theory and Experiment}\ }\textbf {\bibinfo {volume}
  {2019}},\ \bibinfo {pages} {084015} (\bibinfo {year} {2019})}\BibitemShut
  {NoStop}%
\bibitem [{\citenamefont {Bertin}(2010)}]{bertin:10}%
  \BibitemOpen
  \bibfield  {author} {\bibinfo {author} {\bibfnamefont {E.}~\bibnamefont
  {Bertin}},\ }\href {\doibase 10.1088/1751-8113/43/34/345002} {\bibfield
  {journal} {\bibinfo  {journal} {Journal of Physics A: Mathematical and
  Theoretical}\ }\textbf {\bibinfo {volume} {43}},\ \bibinfo {pages} {345002}
  (\bibinfo {year} {2010})},\ \Eprint {http://arxiv.org/abs/arXiv:1004.1597}
  {arXiv:1004.1597} \BibitemShut {NoStop}%
\bibitem [{\citenamefont {Bertin}(2012)}]{bertin:12}%
  \BibitemOpen
  \bibfield  {author} {\bibinfo {author} {\bibfnamefont {E.}~\bibnamefont
  {Bertin}},\ }\href {\doibase 10.1103/PhysRevE.85.042104} {\bibfield
  {journal} {\bibinfo  {journal} {Phys. Rev. E}\ }\textbf {\bibinfo {volume}
  {85}},\ \bibinfo {pages} {042104} (\bibinfo {year} {2012})},\ \Eprint
  {http://arxiv.org/abs/arXiv:1112.0189} {arXiv:1112.0189} \BibitemShut
  {NoStop}%
\bibitem [{\citenamefont {Schr\o{}der}\ \emph {et~al.}(2000)\citenamefont
  {Schr\o{}der}, \citenamefont {Sastry}, \citenamefont {Dyre},\ and\
  \citenamefont {Glotzer}}]{schroder:00}%
  \BibitemOpen
  \bibfield  {author} {\bibinfo {author} {\bibfnamefont {T.~B.}\ \bibnamefont
  {Schr\o{}der}}, \bibinfo {author} {\bibfnamefont {S.}~\bibnamefont {Sastry}},
  \bibinfo {author} {\bibfnamefont {J.~C.}\ \bibnamefont {Dyre}}, \ and\
  \bibinfo {author} {\bibfnamefont {S.~C.}\ \bibnamefont {Glotzer}},\ }\href
  {\doibase 10.1063/1.481621} {\bibfield  {journal} {\bibinfo  {journal} {The
  Journal of Chemical Physics}\ }\textbf {\bibinfo {volume} {112}},\ \bibinfo
  {pages} {9834} (\bibinfo {year} {2000})},\ \Eprint
  {http://arxiv.org/abs/https://doi.org/10.1063/1.481621}
  {https://doi.org/10.1063/1.481621} \BibitemShut {NoStop}%
\bibitem [{\citenamefont {Cavagna}(2009)}]{cavagna:09}%
  \BibitemOpen
  \bibfield  {author} {\bibinfo {author} {\bibfnamefont {A.}~\bibnamefont
  {Cavagna}},\ }\href {\doibase https://doi.org/10.1016/j.physrep.2009.03.003}
  {\bibfield  {journal} {\bibinfo  {journal} {Phys. Rep.}\ }\textbf {\bibinfo
  {volume} {476}},\ \bibinfo {pages} {51 } (\bibinfo {year} {2009})},\ \Eprint
  {http://arxiv.org/abs/arXiv:0903.4264} {arXiv:0903.4264} \BibitemShut
  {NoStop}%
\bibitem [{\citenamefont {Folena}\ \emph {et~al.}(2020)\citenamefont {Folena},
  \citenamefont {Franz},\ and\ \citenamefont {Ricci-Tersenghi}}]{folena:20}%
  \BibitemOpen
  \bibfield  {author} {\bibinfo {author} {\bibfnamefont {G.}~\bibnamefont
  {Folena}}, \bibinfo {author} {\bibfnamefont {S.}~\bibnamefont {Franz}}, \
  and\ \bibinfo {author} {\bibfnamefont {F.}~\bibnamefont {Ricci-Tersenghi}},\
  }\href {\doibase 10.1103/PhysRevX.10.031045} {\bibfield  {journal} {\bibinfo
  {journal} {Phys. Rev. X}\ }\textbf {\bibinfo {volume} {10}},\ \bibinfo
  {pages} {031045} (\bibinfo {year} {2020})},\ \Eprint
  {http://arxiv.org/abs/arXiv:1903.01421} {arXiv:1903.01421} \BibitemShut
  {NoStop}%
\bibitem [{\citenamefont {Folena}\ \emph {et~al.}(2021)\citenamefont {Folena},
  \citenamefont {Franz},\ and\ \citenamefont {Ricci-Tersenghi}}]{folena:21}%
  \BibitemOpen
  \bibfield  {author} {\bibinfo {author} {\bibfnamefont {G.}~\bibnamefont
  {Folena}}, \bibinfo {author} {\bibfnamefont {S.}~\bibnamefont {Franz}}, \
  and\ \bibinfo {author} {\bibfnamefont {F.}~\bibnamefont {Ricci-Tersenghi}},\
  }\href@noop {} {\bibfield  {journal} {\bibinfo  {journal} {Journal of
  Statistical Mechanics: Theory and Experiment}\ }\textbf {\bibinfo {volume}
  {2021}},\ \bibinfo {pages} {033302} (\bibinfo {year} {2021})}\BibitemShut
  {NoStop}%
\bibitem [{\citenamefont {Junier}\ and\ \citenamefont
  {Kurchan}(2004)}]{junier:04}%
  \BibitemOpen
  \bibfield  {author} {\bibinfo {author} {\bibfnamefont {I.}~\bibnamefont
  {Junier}}\ and\ \bibinfo {author} {\bibfnamefont {J.}~\bibnamefont
  {Kurchan}},\ }\href {\doibase https://doi.org/10.1088/0305-4470/37/13/003}
  {\bibfield  {journal} {\bibinfo  {journal} {J. Phys. A}\ }\textbf {\bibinfo
  {volume} {37}},\ \bibinfo {pages} {3945} (\bibinfo {year}
  {2004})}\BibitemShut {NoStop}%
\bibitem [{\citenamefont {Kirkpatrick}\ \emph {et~al.}(1989)\citenamefont
  {Kirkpatrick}, \citenamefont {Thirumalai},\ and\ \citenamefont
  {Wolynes}}]{kirkpatrick:89}%
  \BibitemOpen
  \bibfield  {author} {\bibinfo {author} {\bibfnamefont {T.~R.}\ \bibnamefont
  {Kirkpatrick}}, \bibinfo {author} {\bibfnamefont {D.}~\bibnamefont
  {Thirumalai}}, \ and\ \bibinfo {author} {\bibfnamefont {P.~G.}\ \bibnamefont
  {Wolynes}},\ }\href@noop {} {\bibfield  {journal} {\bibinfo  {journal}
  {Physical Review A}\ }\textbf {\bibinfo {volume} {40}},\ \bibinfo {pages}
  {1045} (\bibinfo {year} {1989})}\BibitemShut {NoStop}%
\bibitem [{\citenamefont {Baity-Jesi}\ and\ \citenamefont
  {Mart\'in-Mayor}(2019)}]{baityjesi:19b}%
  \BibitemOpen
  \bibfield  {author} {\bibinfo {author} {\bibfnamefont {M.}~\bibnamefont
  {Baity-Jesi}}\ and\ \bibinfo {author} {\bibfnamefont {V.}~\bibnamefont
  {Mart\'in-Mayor}},\ }\href@noop {} {\bibfield  {journal} {\bibinfo  {journal}
  {Journal of Statistical Mechanics: Theory and Experiment}\ }\textbf {\bibinfo
  {volume} {2019}},\ \bibinfo {pages} {084016} (\bibinfo {year} {2019})},\
  \Eprint {http://arxiv.org/abs/arXiv:1901.05581} {arXiv:1901.05581}
  \BibitemShut {NoStop}%
\bibitem [{\citenamefont {Metropolis}\ and\ \citenamefont
  {Ulam}(1949)}]{metropolis:49}%
  \BibitemOpen
  \bibfield  {author} {\bibinfo {author} {\bibfnamefont {N.}~\bibnamefont
  {Metropolis}}\ and\ \bibinfo {author} {\bibfnamefont {S.}~\bibnamefont
  {Ulam}},\ }\href@noop {} {\bibfield  {journal} {\bibinfo  {journal} {Journal
  of the American statistical association}\ }\textbf {\bibinfo {volume} {44}},\
  \bibinfo {pages} {335} (\bibinfo {year} {1949})}\BibitemShut {NoStop}%
\bibitem [{\citenamefont {Krauth}(2006)}]{krauth:06}%
  \BibitemOpen
  \bibfield  {author} {\bibinfo {author} {\bibfnamefont {W.}~\bibnamefont
  {Krauth}},\ }\href@noop {} {\emph {\bibinfo {title} {Statistical Mechanics:
  Algorithms and Computations}}}\ (\bibinfo  {publisher} {Oxford University
  Press},\ \bibinfo {address} {Oxford},\ \bibinfo {year} {2006})\BibitemShut
  {NoStop}%
\bibitem [{\citenamefont {Margiotta}\ \emph {et~al.}(2018)\citenamefont
  {Margiotta}, \citenamefont {K\"{u}hn},\ and\ \citenamefont
  {Sollich}}]{margiotta:18}%
  \BibitemOpen
  \bibfield  {author} {\bibinfo {author} {\bibfnamefont {R.~G.}\ \bibnamefont
  {Margiotta}}, \bibinfo {author} {\bibfnamefont {R.}~\bibnamefont {K\"{u}hn}},
  \ and\ \bibinfo {author} {\bibfnamefont {P.}~\bibnamefont {Sollich}},\ }\href
  {http://stacks.iop.org/1751-8121/51/i=29/a=294001} {\bibfield  {journal}
  {\bibinfo  {journal} {Journal of Physics A: Mathematical and Theoretical}\
  }\textbf {\bibinfo {volume} {51}},\ \bibinfo {pages} {294001} (\bibinfo
  {year} {2018})}\BibitemShut {NoStop}%
\end{thebibliography}%

\end{document}